\documentclass{article}
\usepackage{mrs2005,epsfig}
\begin{document} 
\title{DATING SECULAR EVOLUTION: BAR AGES FROM STELLAR KINEMATICS}

\author{D. A. Gadotti (1,2) and Ronaldo E. de Souza (2)}
\affil{(1 - Laboratoire d'Astrophysique de Marseille / France, 2 - Universidade de S\~ao Paulo / Brasil)}

\begin{abstract} 
The last few years witnessed a change of perspective in the context of galaxy formation and evolution. The major role played by bars in the history of these fundamental building blocks of our universe was finally  realized. However, and despite the fact that one of the major concerns of any physical science is to measure timescales for natural phenomena, we have as yet no established way to estimate the ages of these important galactic components. In this contribution we outline the method we developed to estimate bar ages and briefly describe its first results.
\end{abstract} 
 
\section{Why do we need bar ages?}
It is now generally accepted that bars make substantial changes to the destiny of most disk galaxies \cite{kor04}. The development of a method that enables us to measure for how long a bar is present in its host galaxy is thus of primary relevance. Fundamental questions can be better addressed or even answered with such a method at hand, including: To what extent bars are relevant to the building of bulges and its variations \cite{ath05}? How exactly the processes induced by bars contribute to the stellar and non-stellar nuclear activities in disk galaxies \cite{gad01}? When did the first bars appear \cite{elm04}? Are bars a recurrent phenomenon \cite{bou02}? Or are they a robust, perennial galactic structure \cite{she04}?

\section{How to date a bar}
Behind the method we developed in an effort to estimate bar ages lie global dynamical processes. As a bar evolve it buckles out from the plane of the disk it came from to become an structure with an important vertical component (see \cite{gad05} and references therein). These processes raise the vertical velocity dispersion $\sigma_z$ of the stars in the bar. Hence, $\sigma_z$ in a recently formed bar should be similar to that in the disk, while an evolved bar should display values for $\sigma_z$ substantially higher.

We use this strategy to distinguish between young and old stellar bars in a sample of 14 face--on galaxies. Using the 2.3m Bok telescope at Kitt Peak and the 1.5m ESO telescope at La Silla, we obtained spectra along the major and minor axes of these bars and used them to build $\sigma_z$  radial profiles (see Fig. 1). Along the bar major axis our spectra do not reach the end of the bar but reach galactocentric distances where the bar contribution to the collected light is larger than that of the bulge. Thus, at these farther points, the values we measured for the velocity dispersion are a good representation of $\sigma_z$ in the bar and so can be compared with values for $\sigma_z$ in the disk. The latter come from the radial profiles along the bar minor axis. The spectra along that axis cross the bar edge and at the farther points we reliably have measurements of the vertical velocity dispersion in the disk. A comparison between these two values can be used to state whether we are dealing with a young or an evolved bar. In fact, we were able to date all but one of the bars in this sample and show with statistical tests that the two populations thus drawn are different.

With these data we defined two parameters: $\sigma_{z,{\rm bar}}$, that is the stellar vertical velocity dispersion measured in the bar at a galactocentric distance equivalent to $\approx50\%-60\%$ of the bar semimajor axis length; and $\Delta\sigma_z$, that is the difference between $\sigma_{z,{\rm bar}}$ and the stellar vertical velocity dispersion measured in the disk at the same galactocentric distance.
These parameters helped us to identify clear instances of recently formed bars (NGC 1326, NGC 4394, and NGC 5383) and evolved bars (NGC 1302, NGC 1317, and NGC 5850). We show in Fig. 1 a comparison between the $\sigma_z$ radial profiles of young and evolved cases. With these clear cases we found typical values for $\sigma_{z,{\rm bar}}$ in young bars as $\approx30$ km s$^{-1}$ and in old bars as $\approx100$ km s$^{-1}$. Note that this translates to a height scale of about 1.4 Kpc, giving support to scenarios in which bulges may partially form through disk material. Similarly, we found $\Delta\sigma_z\approx5$ km s$^{-1}$ in young bars, in contrast with the value of $\approx40$ km s$^{-1}$ typical for evolved bars.

\section{Results and discussion}
From the 13 barred galaxies in our sample to which we were able to state whether they harbor young or evolved bars, we found that of the eight evolved bars seven (88\%) reside in early--type galaxies (S0--Sa), while only one (12\%) is in an Sb galaxy. On the contrary, the young bars seem to preferentially inhabit later type galaxies: of the five young bars, two (40\%) are found in S0--Sa galaxies and three (60\%) in Sb galaxies.
One also can see that from the eight galaxies with evolved bars, only two (25\%) have AGN, whereas of the five galaxies with young bars, three (60\%) show AGN activity (see also \cite{gad04}). These results must be confirmed by studies with much larger samples since their statistical significance is low. Nevertheless, it is worth noticing that suggestions that bars are young in late--type galaxies can be found in earlier work (see \cite{mar97} and references therein).

Although dating bars can be expensive in terms of telescope time since one needs to explore faint light far from the galactic centers, it is highly desirable to obtain a large number of bar age estimates. On the other hand, the method here outlined is certainly only a first step towards a better understanding on the impacts of bars on the lives of disk galaxies. A better theoretical modelling of the processes that make bars evolve is needed if we want to refine these age estimates.

%
%
\begin{figure}  
\vspace*{1.25cm}  
\begin{center}
\epsfig{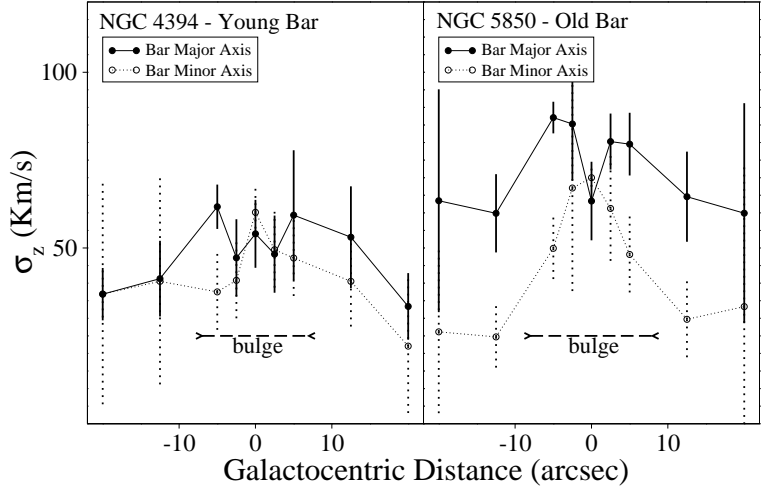}  
\end{center}
\vspace*{0.25cm}  
\caption{Stellar vertical velocity dispersion, $\sigma_z$,
as a function of galactocentric
distance along the bar major and minor axes of two barred galaxies. The information outside
the bulge dominated region is from where one can distinguish recently formed and evolved bars:
along the bar major axis we are still measuring bar kinematics, but outside the bulge, whereas
along the bar minor axis we cross the bar edge and
escape from the bar and bulge light and measure the disk
kinematics. A young bar reveals itself similar to its host disk in what concerns their vertical
extent, as estimated through $\sigma_z$ (left), while an old bar has much higher values for its
stellar vertical velocity dispersion (right). {\it Taken from data in 
\cite{gad05}}
}
\end{figure} 
 
\acknowledgements{Financial support for this work came from FAPESP grants  99/07492-7, 00/06695-0, and 98/10138-8. 
}

\vfill 
\end{document}